\documentclass[aps,prl,reprint,superscriptaddress]{revtex4-1}

\usepackage{graphicx}
\usepackage{dcolumn}
\usepackage{bm}
\usepackage{hyperref}
\usepackage{textcomp}

\begin{document}

\title{Turn-key, high-efficiency Kerr comb source}

\author{Bok Young Kim}
\email{bokyoung.kim@columbia.edu}
\author{Yoshitomo Okawachi}
\author{Jae K. Jang}
\author{Mengjie Yu}
\altaffiliation[Currently at ]{John A. Paulson School of Engineering and Applied Sciences, Harvard University, Cambridge, Massachusetts 02138, USA}
\affiliation{Department of Applied Physics and Applied Mathematics, Columbia University, New York, NY 10027, USA}
\author{Xingchen~Ji}
\affiliation{Department of Electrical Engineering, Columbia University, New York, NY 10027, USA}
\affiliation{School of Electrical and Computer Engineering, Cornell University, Ithaca, NY 14853, USA}
\author{Yun Zhao}
\affiliation{Department of Electrical Engineering, Columbia University, New York, NY 10027, USA}
\author{Chaitanya Joshi}
\affiliation{Department of Applied Physics and Applied Mathematics, Columbia University, New York, NY 10027, USA}
\affiliation{School of Applied and Engineering Physics, Cornell University, Ithaca, NY 14853, USA}
\author{Michal Lipson}
\author{Alexander L. Gaeta}
\affiliation{Department of Applied Physics and Applied Mathematics, Columbia University, New York, NY 10027, USA}
\affiliation{Department of Electrical Engineering, Columbia University, New York, NY 10027, USA}

\date{\today}

\begin{abstract}
  We demonstrate an approach for automated Kerr comb generation in the normal group-velocity dispersion (GVD) regime. Using a coupled-ring geometry in silicon nitride, we precisely control the wavelength location and splitting strength of avoided mode crossings to generate low-noise frequency combs with pump-to-comb conversion efficiencies of up to 41\%, which is the highest reported to date for normal-GVD Kerr combs. Our technique enables on-demand generation of a high-power comb source for applications such as wavelength-division multiplexing in optical communications.
\end{abstract}

\maketitle
Comb generation in microresonators enable integrated and robust platforms \cite{Kippenberg2018, Gaeta2019} that can be used for numerous applications such as spectroscopy \cite{Suh2016, Yu2017, Yu2018, Dutt2018}, distance ranging \cite{Suh2018, Trocha2018}, frequency synthesis \cite{Liang2015, Spencer2018}, optical clocks \cite{Papp2013}, and data communications \cite{Marin-Palomo2017, Fulop2018}. Recently, significant progress has been made in integrating microresonator combs by using low-power, electrically pumped sources \cite{Stern2018, Pavlov2018, Raja2019}, and by utilizing novel generation methods such as pump modulation \cite{Obrzud2017, Cole2018} or thermal control of the resonances \cite{Joshi2016}. These platforms operate in the anomalous group-velocity dispersion (GVD) regime where dissipative cavity solitons are formed \cite{Herr2014}. However, Kerr combs operating in the single-soliton regime have a low pump-to-comb conversion efficiency \cite{Bao2014}. A normal-GVD comb, on the other hand, readily offers high conversion efficiencies and exhibits a slower spectral power falloff \cite{Bao2014, Xue2017}. Normal-GVD combs can be generated through pump modulation at the microresonator free spectral range (FSR) \cite{Lobanov2015} or more commonly by exploiting the coupling between different mode families \cite{Liang2014, Liu2014, Xue2015, Xue2015a, Jang2016, Xue2017}. Although different mode families possess different frequency mode spacings, it is possible for modes in two different families to spectrally overlap and couple to each other leading to a mode splitting at the degeneracy point \cite{Carmon2008, Ramelow2014}. Such an avoided mode crossing effectively creates a local region of anomalous GVD and enables the generation of modulation instability (MI) sidebands \cite{Jang2016}, which is known to be crucial to the initial generation dynamics of a comb \cite{Lamont2013}.

\begin{figure}[!tb]
  \centering
  \includegraphics[width=\linewidth]{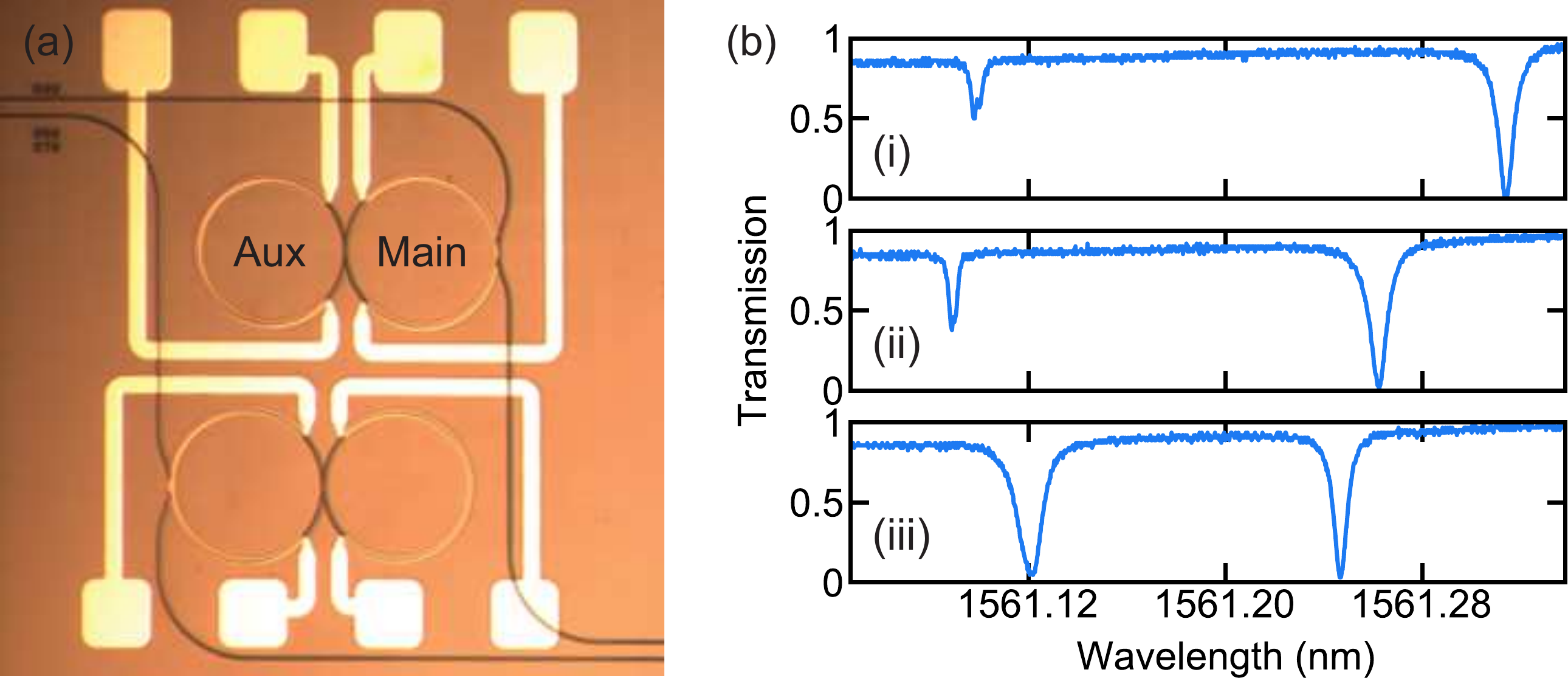}
  \caption{(a) Image of device consisting of coupled ring resonators with integrated platinum resistive heaters. (b) Transmission of device using a counter-propagating probe 1 FSR away from the pump wavelength. Split resonances as the main ring heater is tuned (i),(ii) before parametric oscillation, and (iii) during comb generation.}
  \label{fig:introduction}
\end{figure}

\begin{figure*}[!tb]
  \centering
  \includegraphics[width=\linewidth]{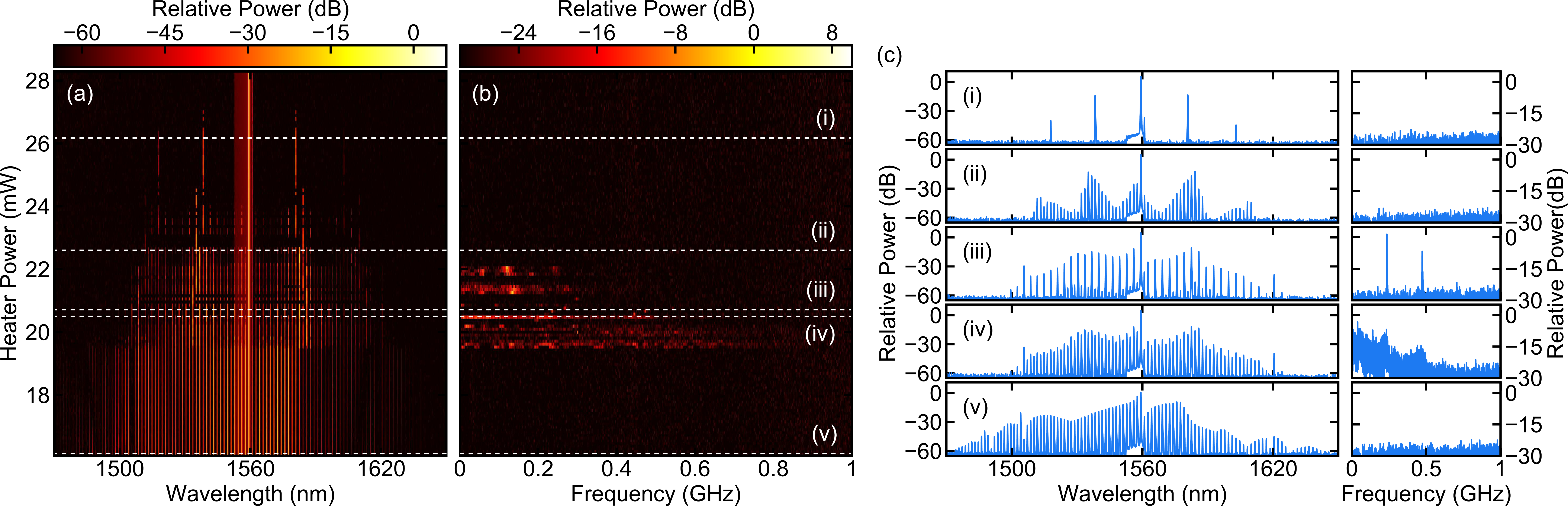}
  \caption{(a) Measured comb spectra and (b) corresponding RF noise spectra for different cavity resonance detunings during the automated process. The power in the main ring heater is tuned from 28 mW to 16 mW while that of the auxiliary ring is set at 82 mW. (c) Optical and RF spectra at heater powers indicated by dashed lines are shown in (a) and (b). The small peak observed 1 FSR away from the pump wavelength is due to the reflection of a counter-propagating probe used to observe the mode splitting during the process. (i) Initial sideband generation from intracavity MI. (ii) Initial filling in of comb lines. (iii) Multi-comb state with sharp noise peaks. (iv) Broadband high-noise comb state. (v) Final low-noise comb state with a high pump-to-comb conversion efficiency.}
  \label{fig:comb_evolution}
\end{figure*}

All previous demonstrations of normal-GVD combs have the disadvantage of requiring tuning the frequency of the pump laser near the mode crossing point. Compared to fixed-frequency lasers, tunable lasers typically have relatively broader linewidths on the order of 100 kHz, which in turn, determines the linewidth of the comb lines. They also have a long-term frequency drift that shifts the center wavelength and limits the stability of the generated comb \cite{Joshi2016}. Depending on device parameters, the required mode interaction (e.g. coupling to higher-order modes) may occur at an undesirable wavelength that could cause non-native comb line spacing \cite{Xue2015} or even inhibit comb generation. To overcome these limitations, a tunable mode interaction should be implemented with a fixed-frequency laser. Comb generation would be delegated to one control variable and allow for straighforward automation of the process, thereby establishing the path to a fully integrated, turn-key microresonator comb system.

In this Letter, we report the first demonstration of microresonator frequency comb generation in the normal-GVD regime using a fixed-frequency laser source. An avoided crossing is introduced using a coupled-ring geometry, and its wavelength location and splitting strength are precisely controlled with integrated heaters. Such control over the mode crossing allows for straightforward use of a computer-controlled system to generate a high-efficiency (41\%) Kerr frequency comb. We show that our system is insensitive to the tuning speed in contrast to soliton generation \cite{Joshi2016}. Finally, we investigate the noise and coherence of the generated comb and determine the conditions for exciting a comb with high coherence.

As shown in Fig. \ref{fig:introduction}\hyperref[fig:introduction]{(a)}, our oxide-clad silicon nitride (SiN) device utilizes coupled ring resonators (main and auxiliary) with integrated platinum resistive heaters. Each microresonator has a waveguide cross section of $730 \times 1000$ nm which allows for the fundamental TE polarization mode to be in the normal-GVD regime near the pump wavelength. An important factor that impacts the comb bandwidth is the strength of the mode coupling. For our normal-GVD comb, an avoided mode crossing is required near the pump wavelength without additional mode interactions elsewhere. Therefore, we exploit the Vernier effect \cite{Miller2015} and design the device to have mode interactions every 50 nm by slightly offsetting the FSRs of each microresonator from one another. The main ring has an FSR of 200 GHz while the auxiliary ring has an FSR of 206 GHz. The coupling gap between the main ring and the auxiliary ring is 525 nm which leads to a mode splitting of 7.1 GHz. To enable high pump-to-comb conversion efficiency, we operate in the over-coupled regime \cite{DelHaye2007} using a bus to main ring gap of 525 nm.

In our experiments, a computer controls an arbitrary waveform generator and individually tunes the resonances of our coupled SiN ring resonators. A narrow linewidth (1 kHz), continuous-wave (CW) fixed-frequency laser at 1559.79 nm [dense wavelength-division multiplexing (DWDM) ITU channel 22] is used as the pump source. The laser light is amplified with an erbium-doped fiber amplifier, and 180 mW of power is coupled into the bus waveguide using a lensed fiber. The free-space output from the device is collimated using an aspheric lens and coupled to a fiber using a collimator. The output is split 80:20 and is collected for analysis by an optical spectrum analyzer (OSA) and an electrical spectrum analyzer, respectively.

Normal-GVD comb generation is initiated by satisfying the phase-matching and power threshold requirements for intracavity MI \cite{Jang2016}. In our system of induced mode interactions, pumping the redshifted resonance at a mode splitting satisfies the phase-matching requirement. A local alteration of the GVD occurs where there is effectively anomalous GVD between the redshifted resonance and the primary MI sideband modes. However, due to the strong interaction between the two microresonators, pumping at the exact crossing results in large loss, and the power threshold requirements cannot be satisfied. To overcome this loss, neighboring resonator modes can be pumped since mode splitting is still present but with reduced loss \cite{Miller2015}. In order to maximize intracavity power while pumping the redshifted resonance at a splitting, the lower wavelength adjacent resonator mode is pumped. Our generation scheme relies on the ability to control the wavelength location and the strength of the mode splitting. By increasing the electrical power through the integrated heaters on our device, the respective microresonators are locally heated and the resonances are redshifted due to the thermo-optic effect \cite{Joshi2016}. Therefore, the mode splitting can be tuned by shifting the frequency degeneracy of the two mode families. Redshifting the smaller (larger) FSR mode family resonances redshifts (blueshifts) the mode crossing. In our method, a main ring resonance is blue detuned towards the pump wavelength $\lambda_p$. However, the blueshifting of the main ring (smaller FSR mode family) resonances is accompanied by the blueshifting of the degeneracy point. To account for this effect, the degeneracy is positioned at a higher wavelength than $\lambda_p$ by a few resonantor modes [\hyperref[fig:introduction]{(i)},\hyperref[fig:introduction]{(ii)} in Fig. \ref{fig:introduction}\hyperref[fig:introduction]{(b)}]. After the heater powers are set accordingly, the main ring resonance is blueshifted towards $\lambda_p$ along with the degeneracy wavelength. Intracavity power builds up leading to a temperature increase in the cavity that offsets the temperature reduction due to the decrease in heater power. Once the intracavity power is large enough, the decrease in heater power does not shift the split resonances [\hyperref[fig:introduction]{(iii)} in Fig. \ref{fig:introduction}\hyperref[fig:introduction]{(b)}].

Figure \ref{fig:comb_evolution} shows the evolution of our normal-GVD comb for part of the computer-controlled process where the electrical power to the main ring heater is tuned from 28 mW to 16 mW while the auxiliary ring is set to a constant 82 mW. As intracavity power begins to grow, primary sidebands due to MI are formed, and the characteristic low radio frequency (RF) noise of the process is observed [\hyperref[fig:comb_evolution]{(i)} in Fig. \ref{fig:comb_evolution}\hyperref[fig:comb_evolution]{(c)}]. The comb then begins to fill in as the resonance is tuned further towards $\lambda_p$ [\hyperref[fig:comb_evolution]{(ii)} in Fig. \ref{fig:comb_evolution}\hyperref[fig:comb_evolution]{(c)}]. The primary sidebands then grow and serve as pump fields that generate separate combs and induce sharp peaks in the RF spectrum [\hyperref[fig:comb_evolution]{(iii)} in Fig. \ref{fig:comb_evolution}\hyperref[fig:comb_evolution]{(c)}]. Tuning further into the resonance, we see a broadband high-noise state [\hyperref[fig:comb_evolution]{(iv)} in Fig. \ref{fig:comb_evolution}\hyperref[fig:comb_evolution]{(c)}]. The rise in RF noise is attributed to the broadband MI gain from the two strong primary sidebands serving as pump fields \cite{Okawachi2015}. As the power in the cavity increases, noise is generated from spontaneous FWM. Finally, as the resonance is further blue detuned, we observe a sharp drop in noise [\hyperref[fig:comb_evolution]{(v)} in Fig. \ref{fig:comb_evolution}\hyperref[fig:comb_evolution]{(c)}], and the comb is phase-locked while maintaining a high pump-to-comb conversion efficiency. A key reason our method for comb generation can be automated is that the desired comb state persists over a large heater tuning range from 20 mW to 16 mW [Fig. \ref{fig:comb_evolution}\hyperref[fig:comb_evolution]{(a)}]. In contrast to conditions required for the soliton state, where the heater scan speed must be close to the thermal time constant of the ring to counter the thermal recoil \cite{Joshi2016}, the large detuning range in our system allows for arbitrary heater scan speeds to consistently reach the low-noise comb state.

\begin{figure}[!tb]
  \centering
  \includegraphics[width=0.9\linewidth]{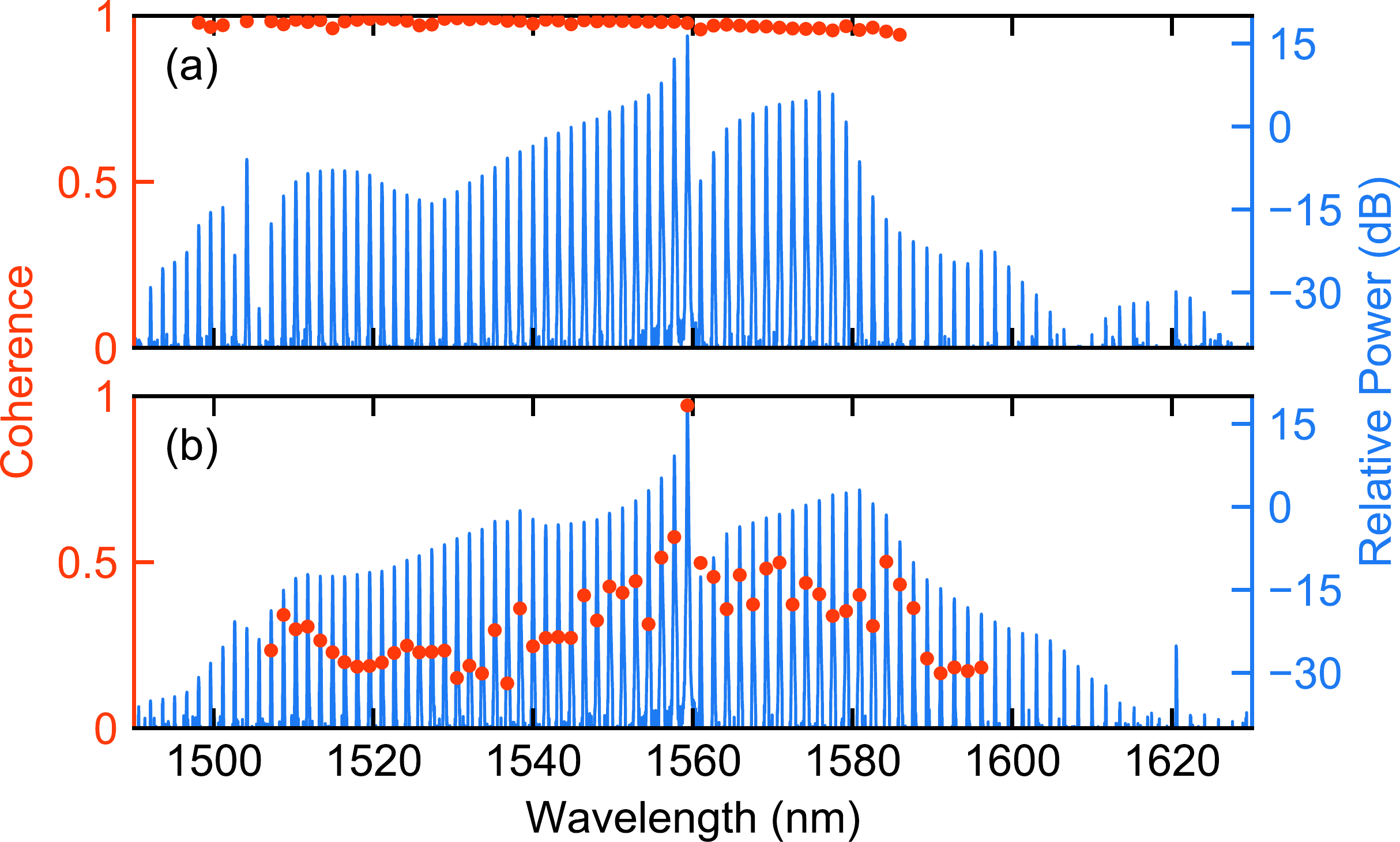}
  \caption{Coherence measurement of (a) low-noise comb state with high pump-to-comb conversion efficiency, and (b) the high-noise comb state.}
  \label{fig:coherence}
\end{figure}

\begin{figure}[!tb]
  \centering
  \includegraphics[width=0.9\linewidth]{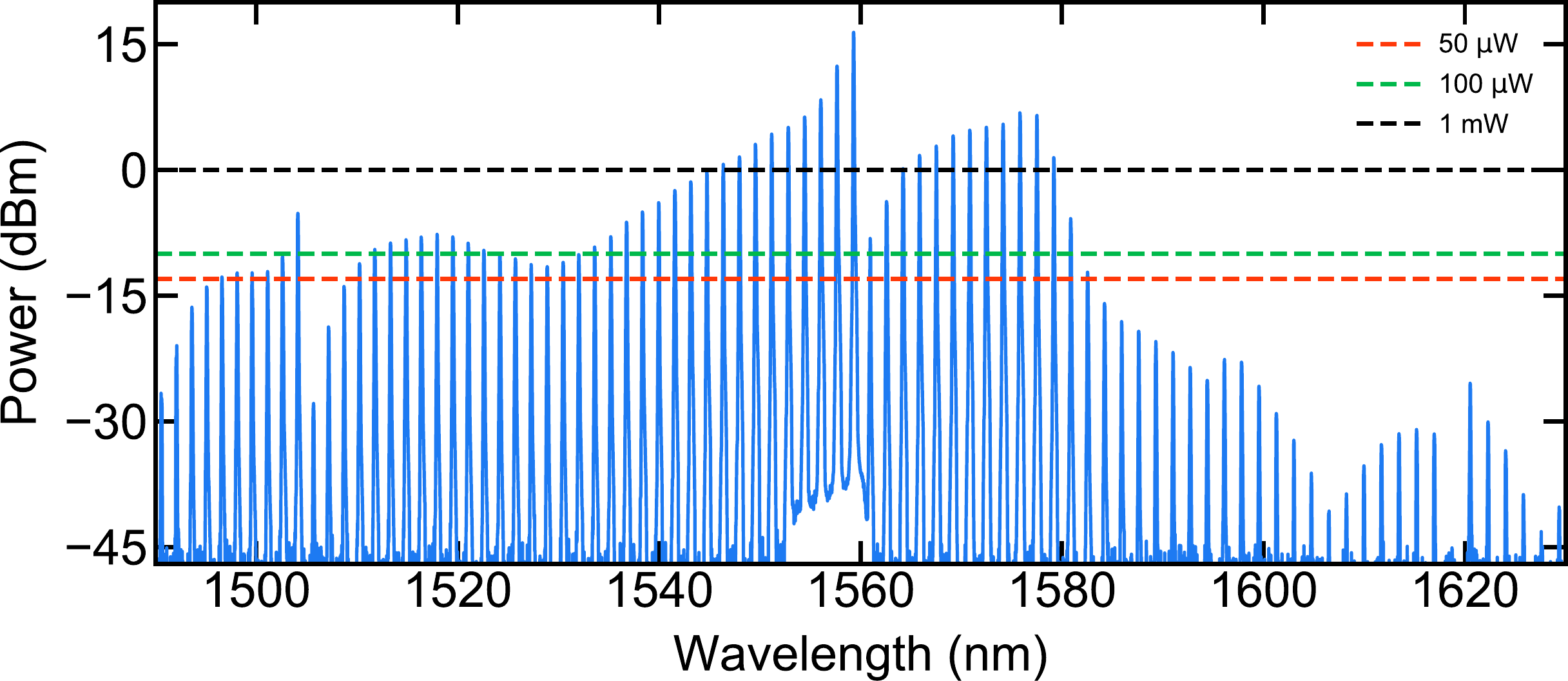}
  \caption{Measured comb spectrum from automated generation with a comb line spacing of 201.6 GHz and a 41\% pump-to-comb conversion efficiency. Power levels of 50 \textmu W, 100 \textmu W, and 1 mW are given by red, green, and black dashed lines respectively. There are 18 lines with powers greater than 1 mW, 38 lines with powers greater than 100 \textmu W, and 51 lines with powers greater than 50 \textmu W.}
  \label{fig:efficiency}
\end{figure}

The RF noise is directly correlated to the coherence of the comb. We measure the coherence by using the technique given in \cite{Webb2016}. For this setup, we use a fiber-based asymmetric Mach-Zehnder interferometer to overlap the output sampled at two different times and create spectral fringes from their relative time delay. The path-length difference is set to 3 m, which corresponds to a few photon lifetimes and a linewidth resolution of 100 MHz. Such a laser laser linewidth is sufficient for digital modulation applications such as differential phase-shift keying. Utilizing the instability of the interferometer, we obtain 50 OSA traces to measure the visibility of these spectral fringes. The visibility at wavelength $\lambda$ is given by
\begin{equation}
  V(\lambda) = \frac{I_{\mathrm{max}}(\lambda) - I_{\mathrm{min}}(\lambda)}{I_{\mathrm{max}}(\lambda) + I_{\mathrm{min}}(\lambda)},
\end{equation}
where $I_{\mathrm{max}}(\lambda)$ and $I_{\mathrm{min}}(\lambda)$ are the maximum and minimum intensities at $\lambda$, respectively, and the resulting coherence is
\begin{equation}
  \left|g_{12}^{(1)}(\lambda)\right| = \frac{I_1(\lambda) + I_2(\lambda)}{2\sqrt{I_1(\lambda)I_2(\lambda)}}V(\lambda),
\end{equation}
where $I_1(\lambda)$ and $I_2(\lambda)$ are the intensities in each arm of the interferometer \cite{Gu2003}. After calibrating for power differences in the arms, we show that the relevant lines of the generated comb are highly coherent with a minimum coherence of 94\% [Fig. \ref{fig:coherence}\hyperref[fig:coherence]{(a)}]. We also measure the coherence of the comb in a high-noise state. Figure \ref{fig:coherence}\hyperref[fig:coherence]{(b)} shows the calculated coherence for the high-noise comb state, and we observe, as expected, that aside from the pump line the comb lines show low coherence.

We examine the pump-to-comb conversion efficiency, which is defined to be the ratio between the power of all the comb lines sans the pump line and the pump power in the bus waveguide. For the generated spectrum shown in Fig. \ref{fig:efficiency}, the comb line spacing is 201.6 GHz, and the conversion efficiency is measured to be 40.6\%. Figure \ref{fig:efficiency} shows 18 lines with powers greater than 1 mW, 38 lines with powers greater than 100 \textmu W, and 51 lines with powers greater than 50 \textmu W. The 51 comb lines occur sequentially in mode number between 1495 nm and 1583 nm except for the 3 modes lying between 1505 nm and 1509 nm. These 3 comb lines are weaker in power due to the periodicity of the induced avoided mode crossing. Similar depletion of comb line powers can be observed 1 higher comb mode from the pump and near 1619 nm. Therefore, increasing the frequency separation between mode interactions would eliminate such depletions of comb line powers and potentially broaden the comb. The comb bandwidth can also be increased by reducing the GVD through dispersion engineering and by increasing the mode splitting \cite{Jang2016}. In the normal-GVD regime, the FSR of a resonator decreases with frequency. As a result, the phase-matched MI process will occur at more distant modes if the magnitude in FSR change is reduced through the reduction in GVD. Likewise, increasing the mode splitting will allow for greater compensation of the GVD and broaden the comb.

\begin{figure}[!tb]
  \centering
  \includegraphics[width=0.9\linewidth]{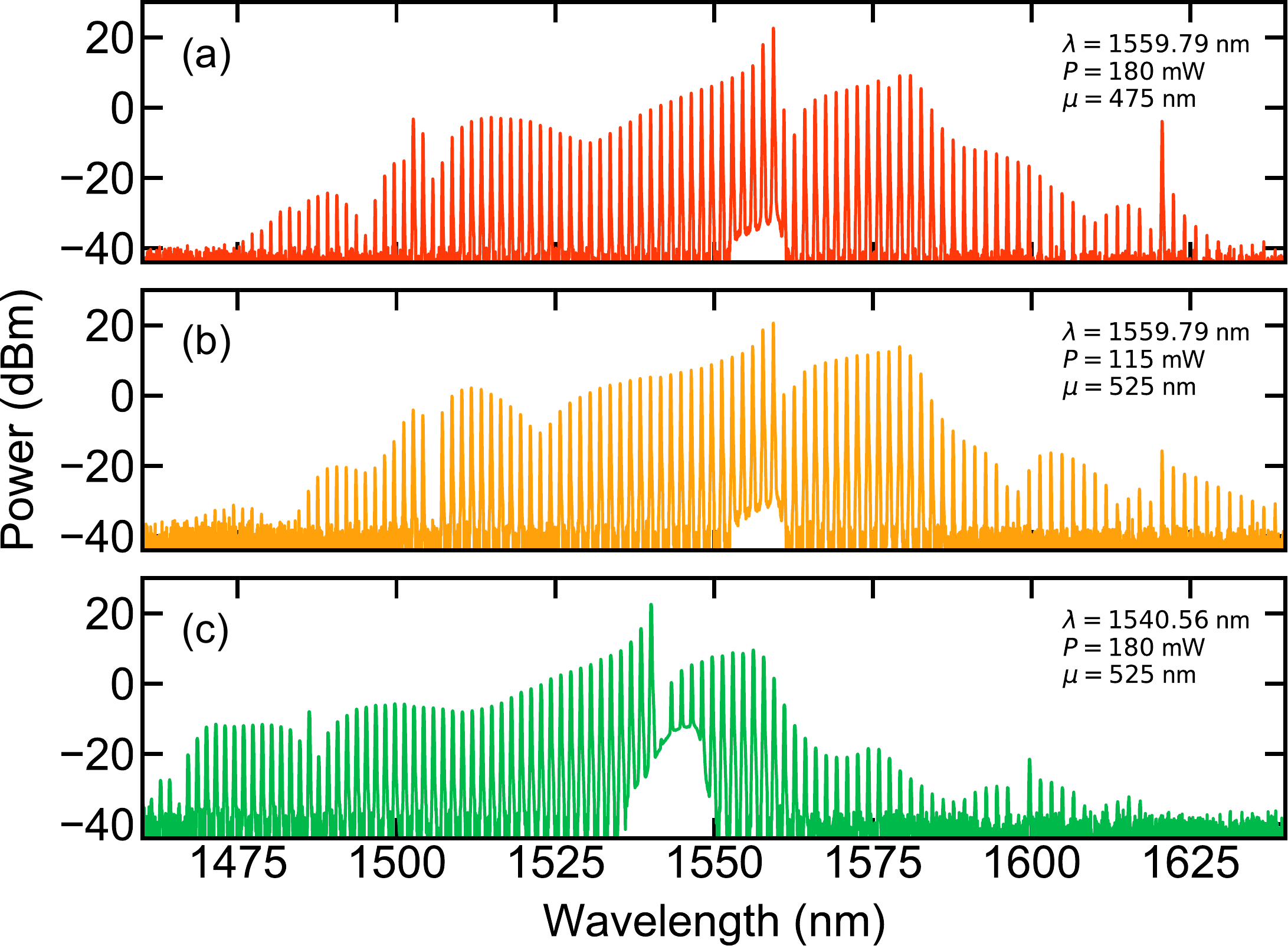}
  \caption{Measured comb spectra from automated generation in different configurations for the coupled-ring geometry. (a),(b) Pump wavelength set at 1559.79 nm. (a) Pump power set at 180 mW with 475-nm coupling gap between main and auxiliary ring. (b) Pump power set at 115 mW with 525-nm coupling gap. (c) Pump wavelength set at 1540.56 nm and pump power set at 180 mW with 525-nm coupling gap.}
  \label{fig:other}
\end{figure}

Lastly, we show that the automated method applies generally across other configurations for the coupled-ring geometry. Figure \ref{fig:other}\hyperref[fig:other]{(a)} shows the spectrum when the coupling gap between the rings is 475 nm, which corresponds to a stronger mode interaction. Other devices were also tested that generated similar spectra indicating this method is not limited to one device. Returning to the original device, Fig. \ref{fig:other}\hyperref[fig:other]{(b)} shows the spectrum when the pump was set to 115 mW. Similar spectra were generated for a range of pump powers from 115 mW to 200 mW. Figure \ref{fig:other}\hyperref[fig:other]{(c)} shows the spectrum when the pump wavelength is set at 1540.56 nm (DWDM ITU channel 46) by a narrow linewidth (1 kHz), CW fixed-frequency laser. We have also used a tunable CW laser to set different pump wavelengths and successfully generated the final comb state.

In conclusion, we have demonstrated a systematic approach to generating high-power Kerr combs using a fixed-frequency laser source. Our technique increases the stability of a comb and can overcome variations in the parameter settings arising from minute differences in fabrication. The on-demand generation method produces low-noise and high-power comb lines that are useful for applications such as communications by eliminating the need for multiple laser sources. Furthermore, high-power combs are useful in other areas such as spectroscopy where the detection of power in each comb line is critical. By automating the generation scheme, we have shown that turn-key operation is a reality for such microresonator frequency comb systems.
\newline
\newline
\noindent\textbf{Funding.} Advanced Research Projects Agency-Energy (ARPA-E) (\text{DE-AR0000843}); Air Force Office of Scientific Research (AFOSR) (FA9550-15-1-0303); National Science Foundation (NSF) (ECS-0335765); US Department of Defense (DOD) under the Air Force Small Business Innovation Research (SBIR) (FA8650-19-C-1002).
\newline
\newline
\noindent\textbf{Acknowledgment.} This work was performed in part at the Cornell Nano-Scale Facility, a member of the National Nanotechnology Infrastructure Network, which is supported by the NSF.

\bibliography{paper}

\end{document}